\documentclass[aps,pra,twocolumn]{revtex4-1}

\usepackage{amsmath}
\usepackage{amssymb}

\begin{document}


\title{A Robust Mathematical Model for Loophole-Free Clauser-Horne Experiments}


\author{Peter Bierhorst}
\email[pbierhor@tulane.edu]\homepage{}
\affiliation{Tulane University}

\date{August 1, 2014}

\begin{abstract}
Recent experiments \cite{giustina:2013,christensen:2013} have reached detection efficiencies sufficient to close the detection loophole, testing the Clauser-Horne (CH) version of Bell's inequality. For a similar future experiment to be completely loophole-free, it will be important to have discrete experimental trials with randomized measurement settings for each trial, and the statistical analysis should not overlook the possibility of a local state varying over time with possible dependence on earlier trials (the ``memory loophole"). In this paper, a mathematical model for such a CH experiment is presented, and a method for statistical analysis that is robust to memory effects is introduced. Additionally, a new method for calculating exact p-values for martingale-based statistics is described; previously, only non-sharp upper bounds derived from the Azuma-Hoeffding inequality have been available for such statistics. This improvement decreases the required number of experimental trials to demonstrate non-locality. The statistical techniques are applied to the data of \cite{giustina:2013} and \cite{christensen:2013} and found to perform well.
\end{abstract}


\maketitle


\section{Introduction}\label{s:introduction}

The relationship between quantum mechanical predictions and the principle of locality was examined by Einstein, Podolsky, and Rosen in \cite{EPR}. In this paper, it was suggested that an underlying refinement of quantum mechanics, compatible with locality, could possibly provide a fuller description of nature. Such a refinement has come to be known as a ``local hidden variable theory." It was later shown by Bell \cite{BELL} that there are experimental configurations in which quantum mechanics makes predictions that cannot be explained by \emph{any} local hidden variable theory. 

Such an experiment is known as a Bell test, and to date, no Bell test experiment has been executed with sufficient precision to successfully rule out all possible local hidden variable theories. In experiments with entangled photons, one persistent issue has been the ``detection loophole," whereby if too many systems are undetected, a local hidden variable theory can mimic quantum correlations by occasionally evading measurement. However, the detection loophole can be eliminated if detector efficiencies exceed a certain critical threshold \cite{eberhard:1993}. Promisingly, two recent experiments \cite{giustina:2013,christensen:2013} have been able to exceed this threshold in a photon experiment, taking an important step towards the ultimate goal of a fully loophole-free Bell test. However, neither of the two experiments \cite{giustina:2013,christensen:2013} enforced space-like separation between detection events, and so a local hidden variable theory explanation for the violation of the Bell inequality cannot be definitively ruled out. 

The papers \cite{giustina:2013,christensen:2013} make use of Clauser-Horne (CH) type Bell inequalities \cite{CH74}, which are constraints on the probabilities of various experimental outcomes that must be obeyed by local theories. To analyze the experimental results, the papers \cite{giustina:2013,christensen:2013} calculate sample standard deviations of the data and find that the inequalities are violated by 70 standard deviations for \cite{giustina:2013} and 7 for \cite{christensen:2013}. However, this method of analysis is known to have flaws \cite{GILL,barrett:2002,zhang:2011}. For instance, the sample standard deviation can only represent a valid estimate of a physical parameter if the photon states being tested over successive trials are assumed to be independent and identically distributed (i.i.d.). But the possibility that the state of the system varies in time, with possible dependence on the outcomes of earlier trials, is allowed by local hidden variable theories. This issue is known as the ``memory loophole," and while it was shown by Gill \cite{GILL} that local hidden variable theories cannot exploit the memory loophole to violate the CHSH inequality \cite{CHSH}, the issue does necessitate a different form of statistical analysis. Such an analysis has not yet been adapted to the scenario of a CH-type experiment.

The necessity of various experimental parameters could also use clarification. For instance, the importance of having discrete experimental trials, to avoid the coincidence-time-loophole \cite{larsson:2004}, was recognized in \cite{christensen:2013}. But it is also important to have randomized measurement settings prior to each trial.

It would be good to address these issues prior to the design of a future loophole-free CH experiment. This paper puts the analysis of CH experiments on solid theoretical ground, clearly delineating a sufficient set of experimental parameters for a loophole-free test, and demonstrating how to perform a correct statistical analysis of the results that does not ignore possible memory effects. The statistical analysis is based only on the same minimal set of assumptions used to derive the Bell inequalities, and does not require additional assumptions such as independent, identically distributed experimental trials.

The statistical analysis of experimental results is formulated as a classic hypothesis test. In a hypothesis test, one analyzes a \emph{test statistic} and calculates the corresponding \emph{p-value} to quantify the evidence for rejecting the null hypothesis. A test statistic is a real-valued function of the experimental data, and for a given null hypothesis, one can predict that the test statistic should fall within a certain range with high probability. If the experimentally-calculated value of the test statistic deviates from this expected range, one may decide to reject the null hypothesis. Then to quantify the evidence against the null hypothesis, one calculates the p-value: the probability of seeing a test statistic as or more extreme than the experimentally observed value, assuming the null hypothesis is true. In our setting, the null hypothesis is that nature obeys a local hidden variable theory, so small p-values will correspond to evidence against this hypothesis.

The test statistics employed in this paper are \emph{supermartingales}. (A supermartingale is a type of random process; supermartingales arise naturally in the presence of memory effects.) It is straightforward to assign a p-value to a binary-output supermartingale, but some of the statistics will take an output in a larger finite set. In this paper, we introduce a novel method that allows for the calculation of exact p-values for all finite-output supermartingale test statistics; previous analyses \cite{GILL,zhang:2013} have only provided non-sharp upper bounds for such p-values by appealing to the Azuma-Hoeffding inequality \cite{HOEFFDING,AZUMA}. With this improvement in statistical strength, fewer experimental trials are necessary to see statistical significance, which could be quite valuable for an experiment in which components wear out quickly.

The structure of the paper is as follows: in Section \ref{s:designing}, we describe the experimental set-up for a loophole-free test of a CH-type Bell inequality and introduce a mathematical model for the experiment that allows us to define a useful test statistic. In Section \ref{s:statistics}, we show how to perform a hypothesis test and analyze the test statistic without making any tacit assumptions. Other candidate test statistics are also explored, and a new method for finding exact p-values for finite-output martingale statistics is also described. Finally, the statistical techniques are applied to the data of \cite{giustina:2013} and \cite{christensen:2013} as a test case and found to perform well.  The appendix contains a deeper discussion of the potential loopholes present in CH-type experiments, a technical proof of one of the claims in the paper, a classification of all relevant CH-type inequalities, and a generalization of the techniques of the paper to the scenario in which the measurement setting probabilities are not required to be equal.

\section{Designing a Loophole-Free Experiment}\label{s:designing} 

\subsection{Experimental Basics}\label{s:basics}

CH-type inequalities apply to two-detector, two-setting, two-outcome ($2\times 2\times 2$) experiments. In such experiments, there are two detector apparatuses in two separate locations, labeled location ``1" and ``2," and each detector can be set in one of two configurations. The two configurations, or ``measurement settings," of the first detector are labeled $a$ and $a'$, and the measurement settings of the second detector are labeled $b$ and $b'$. No matter the setting, the outcome at each detector always comes from a two element set, labeled $\{+,0\}$. This is the basic set-up for a CH-type experiment

We also need to explicitly lay out a set of experimental requirements that, if met, will be sufficient for a loophole-free test of a CH-type inequality. The first requirement is that the experiment be performed as a sequence of discrete trials. By doing this, the experimenter can rule out the coincidence-time loophole, which was described in \cite{larsson:2004} and cited in \cite{christensen:2013} as a motivating factor for choosing to set up an experiment with discrete trials. There may be other ways to get around the coincidence-time loophole by carefully post-processing experimental data \cite{larsson:2013}, but the most natural solution is to have the discrete trials in place as part of the experimental design.

We also require that the experimental settings are randomized right before each experimental trial, in such a way that the event corresponding to the random choice of the setting at detector 1 for the $i$th trial is space-like separated from the event corresponding to the end of the $i$th trial at detector 2, and vice-versa. If the space-like separations are not in place, it cannot be ruled out that Bell-inequality violations could be achieved by something as mundane as sub-luminal signaling. The randomization of the detector settings is motivated by the ``free will" assumption, which is used in the derivation of Bell inequalities. Mathematically, ``free will" states that the random process governing the selection of the measurement settings is independent of the state ``$\lambda$" of the system of the approaching photons. Accordingly, care should be taken to ensure that the random process governing the measurement setting is indeed unrelated to the photon source, to the best extent that it is possible to do this. 

Finally, we also assume that all four measurement setting configurations are equally likely. This is a common experimental practice and it simplifies the mathematical analysis. This assumption is not strictly necessary for a loophole-free test of a CH-type inequality, and we explore some of the consequences of relaxing the assumption in Appendix \ref{s:mod}. 

The above experimental conditions are sufficient for a loophole-free test of a CH-type Bell inequality. Slight modifications these conditions could theoretically still be sufficient for a loophole-free experiment, but the above set is quite natural, and perhaps optimal given practical constraints imposed by current equipment. A discussion of experimental loopholes in Appendix \ref{s:loopholes} provides some additional motivation for this claim.

\subsection{A Mathematical Model and a Test Statistic}

Now that we have laid out the experimental basics, the next step is to construct a mathematical model for the experiment. The first two columns of Table \ref{t:trialsubset} show how the first 11 trial results of a CH-type experiment might look. (We will explain the meaning of the other columns of Table \ref{t:trialsubset} later.) To explain the notation in this table, the trial outcome ``0+$a'b$" should read ``detector 1 outputs 0 while in setting configuration $a'$, detector 2 outputs + while in setting configuration $b$." As there are two possible choices for each of the four components of a trial outcome, there are sixteen total possible outcomes. We define a sequence of random variables $\{C_i\}$ to be the outcomes of the trials:
\begin{equation*}
C_i = \text{ the outcome of the $i$th trial.}
\end{equation*}

If the experiment is described by a local hidden variable theory, then probabilities of various outcomes for the $C_i$ random variables will satisfy linear constraints known as Bell inequalities. In a realistic CH-type experiment with photons, most trials will be ``00" outcomes that will not exhibit non-locality. (The reason for this is that the power of the photon source must be turned down very low to ensure that when photon states are produced, there is a very small probability of creating a multi-photon state that may not violate the chosen Bell inequality. This necessarily results in a high proportion of 00 counts. A detailed analysis of relevant photon sources can be found in \cite{vivoli:2014}.) The most useful Bell inequalities will thus be constraints that do not refer to these 00 trials. The original CH 74 inequality \cite{CH74} is a $2\times 2\times 2$ inequality that does not involve the ``00" outcomes, and can be adapted to our situation, as will be demonstrated in Section \ref{s:statistics}. However, the statistic $J$ introduced by Eberhard in \cite{eberhard:1993} and analyzed by \cite{giustina:2013} is a more natural fit for our experimental set-up.

\begin{table}[h]\caption{A possible set of outcomes for the first few trials of a CH-type experiment.}\label{t:trialsubset}
\begin{center}
\begin{tabular}{ c|c|c|c|c|c| }
Trial & Outcome & $C_i$ & $J'_i$ & $J_k$ & $S_j$ \\
 \hline
1 & 00$ab$ & $C_1$ & $J'_1=0$ &  & \\
2 & 00$ab'$ & $C_2$ & $J'_2=0$ & & \\
3 & 0+$a'b$ & $C_3$ & $J'_3=-1$ & $J_1=-1$ & $S_1$\\
4 & 00$ab$ & $C_4$ & $J'_4=0$ & &\\
5 & 00$a'b$ & $C_5$ & $J'_5=0$ & &\\
6 & ++$a'b'$ & $C_6$ & $J'_6=-1$ & $J_2=-1$ & $S_2$\\
7 & 00$a'b$ & $C_7$ & $J'_7=0$ & &\\
8 & 0+$ab'$ & $C_8$ & $J'_8=0$ & & $S_3$\\
9 & 00$ab'$ & $C_9$ & $J'_9=0$ & & \\
10 & ++$ab$ & $C_{10}$ & $J'_{10}=+1$ & $J_3=+1$ & $S_4$ \\
11 & 00$a'b'$ & $C_{11}$ & $J'_{11}=0$ &  & \\
\vdots & \vdots & \vdots & \vdots & \vdots &\vdots \\

 \end{tabular}
\end{center}
\end{table}

The original Eberhard statistic allows for three outcomes, but if two of the outcomes are lumped together into one, as was done in \cite{giustina:2013}, the statistic can be applied to our current scenario. The argument of \cite{eberhard:1993} (as adapted by \cite{giustina:2013}) tells us that if the measurement settings are equiprobable, then under a local hidden variable theory the total number of ++$ab$ counts must be exceeded by the total number of the sum of all +0$ab'$, 0+$a'b$, and ++$a'b'$ counts, in expectation. However, the argument in \cite{eberhard:1993} relies on the tacit assumption that the outcome probability distribution cannot change in response to the results of earlier trials, and \cite{eberhard:1993} offers only an intuitive assessment of a threshold for statistical significance (``let us say, ten standard deviations"). Our goal is to assign a precise p-value to a deviation away from the local realm, without any tacit assumptions (such as, for instance, i.i.d trials). To do this, we start by re-formulating this Bell constraint in terms of $C_i$:
\begin{multline}\label{e:eberhardo}
P(C_i=\text{++}ab)\le\\
P(C_i=\text{++}ab) + P(C_i=\text{+0}ab') + P(C_i=\text{0+}a'b).
\end{multline}
The above can be proved with standard Bell-inequality-style arguments, but (\ref{e:eberhardo}) is not sufficient to properly analyze accumulated tallies over many trials. 

To analyze the tallying process over multiple trials, we can define a statistic
\begin{equation}\label{e:Jrawdef}
J'_i = \begin{cases} +1, & \text{if $C_i=$++$ab$ } \\ 
 -1, & \text{if $C_i=$+0$ab'$, 0+$a'b$, or ++$a'b'$,}\\
0, & \text{if $C_i$ takes on any other value},
\end{cases}
\end{equation}
and then it is clear that under a local hidden variable theory,
\begin{equation}\label{e:use1}
P(J'_i=+1)\le P(J'_i=-1).
\end{equation}
The statistic $\sum_{i=1}^n J'_i$ can be used as a measure of the accumulated violation of the Bell inequality (\ref{e:eberhardo}) after $n$ trials. However, to do this without tacit assumptions actually requires the following modified version of (\ref{e:use1}):
\begin{equation}\label{e:use2}
P(J'_i=+1|\mathcal{J}'_i)\le P(J'_i=-1|\mathcal{J}'_i),
\end{equation}
where $\mathcal{J}'_i$ is allowed to be any event consisting of outcomes of the earlier trials -- that is, outcomes of random variables in the set $\{J'_k\}_{k=1}^{i-1}$. In an intuitive sense, expression (\ref{e:use2}) states that under a local hidden variable theory, knowledge of the outcomes of earlier trials cannot be used to increase the probability that $J'_i$ equals +1 any higher than 50\%. Expression (\ref{e:use2}) is closely related to the statement that the sequence $\{J'_i\}$ is a \emph{supermartingale} -- that is, for any $n\ge 1$, the following conditions hold:
\begin{eqnarray*}
&E\left(\left|\sum_{i=1}^n J'_i\right|\right)<\infty&\\ 
&E(\sum_{i=1}^{n+1} J'_i \mid J'_1,...,J'_n)\le \sum_{i=1}^{n} J'_i,&
\end{eqnarray*}
where $E(\cdot )$ denotes the expectation of a random variable. Under a local hidden variable theory, expression (\ref{e:use2}) holds for any choice of $\mathcal{J}'_i$ for which $P(\mathcal{J}'_i)>0$. This can be proved by arguments such as the ones found in \cite{GILL} or \cite{bierhorst1:2013}. The intuitive reason for why (\ref{e:use2}) holds is the fact that the settings are re-randomized prior to each trial, and knowledge of past outcomes does not help the local hidden variable predict the measurement settings for the next trial, which is really what would be necessary to ``beat" the constraint (\ref{e:eberhardo}).

Expression (\ref{e:use2}) tells us that under a local hidden variable theory, the sequence $J'_1, J'_2,...$ cannot accumulate a preponderance of ``+1" tallies to ``$-1$" tallies any better than a fair coin can accumulate many ``heads" to relatively few ``tails." Unfortunately, $\sum_{i=1}^n J'_i$ is a poor choice of statistic for measuring the deviation in the positive direction. For a real-world CH experiment such as \cite{christensen:2013}, something in the order of 99.5\% of trials will output ``00", so accordingly, at least 99.5\% of the $J'_i$ outcomes will be zero. Among the less than .5\% of the $J'_i$ assuming values in the set $\{-1,+1\}$, Quantum Mechanics predicts that the +1 values would slightly predominate. However, a local hidden variable theory obeying (\ref{e:use2}) could simulate $J'_i$ behavior where $J'_i$ \emph{always} takes a nonzero value, with a 50-50 distribution on $-1$ and $+1$. After, say, $n=2,000,000$ trials, the standard deviation of $\sum_{i=1}^n J'_i$ under this local hidden variable theory is about 700, with an expected value of zero. Unfortunately, the expected value of $\sum_{i=1}^n J'_i$ under the quantum theory may have only crept up to a couple hundred by this point, in which case the p-value for the extremity of $\sum_{i=1}^n J'_i$ would be only slightly less than $\frac{1}{2}$. (Note that this particular issue is germane to the CH scenario: a supermartingale statistic based on the CHSH inequality, such as the one studied in \cite{GILL}, always assumes a value in the binary set $\{-1,+1\}$ for any trial outcome.)

It is possible to get around this bug by going out to very large $n$, because the quantum expectation for $\sum_{i=1}^n J'_i$ grows linearly with $n$, whereas the standard deviation of $\sum_{i=1}^n J'_i$ under the local hidden variable theory described in the previous paragraph grows linearly with $\sqrt n$. However, there is a more efficient approach. The solution is to condition on a nonzero outcome for $J'_i$, and thus analyze only the experimental trials where the sum of the $J'_i$ changes value. 

To do this, we define a new sequence of statistics $\{J_k\}$ within our model. Let $E$ be the 4-element subset of outcomes of $C_i$ that appear in (\ref{e:eberhardo}) -- that is, $E=\{\text{++}ab,\text{+0}ab',\text{0+}a'b,\text{++}a'b'\}$. Then define $T_k$ to be the $k$th $C_i$ that takes a value in $E$ and define
\begin{equation}\label{e:Jdef}
J_k = \begin{cases} +1, & \text{if $T_k=$++$ab$ } \\ 
 -1, & \text{if $T_k=$+0$ab'$, 0+$a'b$, or ++$a'b'$.}  \end{cases}
\end{equation}
Mathematically, it is of course possible that $T_k$ is undefined, which will happen if fewer than $k$ of the $C_i$ take a value in $E$. In this degenerate case, we can define $J_k$ to be $-1$, a useful trick for certain formal arguments appearing in Appendix \ref{s:subsequence}. This technicality will not come in to play in the analysis of experimental data, because conclusions should only be drawn based on values of $k$ for which at least $k$ of the $C_i$ have taken a value in the set $E$. 

Table \ref{t:trialsubset} has a column illustrating how the values of the various $J_k$ will be set by the experimental data. Importantly, under a local hidden variable theory, the sequence $J_k$ is a supermartingale, satisfying an expression similar to (\ref{e:use2}):
\begin{equation}\label{e:use3}
P(J_k=+1|\mathcal{J}_k)\le P(J_k=-1|\mathcal{J}_k),
\end{equation}
where $\mathcal{J}_k$ is any positive-probability event consisting of outcomes of $J_i$ variables for $i<k$. The proof that (\ref{e:use3}) holds is given in Appendix \ref{s:subsequence}. (\ref{e:use3}) implies that the statistic $J
:=\sum_{k=1}^m J_k$, where $m$ is the total number of trials that take a value in the set $E$, can be used as a measure of the accumulated violation of the Bell inequality (\ref{e:eberhardo}). (Note that the value of $J$ defined this way coincides with the value of $J$ as defined in \cite{eberhard:1993} and \cite{giustina:2013}, except for multiplication by a factor of $-1$.) We now demonstrate how to assign a precise p-value to the extremity of $J$.

\section{Statistical Considerations and the Memory Loophole}\label{s:statistics}

\subsection{Eberhard-style statistics}

For a CH experiment like \cite{giustina:2013,christensen:2013}, the statistic $J$ should accumulate a Bell violation over many trials. We can see this by forming empirical distributions for the non-00 outcomes from the data of these experiments, which are given in Table \ref{t:empiricaldistributions} \footnote{To form Table \ref{t:empiricaldistributions}, the data for \cite{giustina:2013} was obtained from the follow-up note \cite{kofler:2013}. This data has not been filtered into discrete time intervals, as is necessary for proper analysis (see \cite{larsson:2013}); presumably the empirical distribution would exhibit non-locality by a smaller amount after this step is taken. As for the calculation of the distribution based on \cite{christensen:2013}, a normalization step was required to take into account the small but significant deviation from an equiprobable setting distribution. The apparent deviation from equiprobability for measurement settings in the data of \cite{christensen:2013} does not result from a mis-calibration of the random process that governed the measurement settings. There were 4,450 random re-assignments of the detector settings, and about 25,000 trials were performed for each assignment of the settings. The 4,450 setting blocks were close to evenly split among the four setting configurations, with a small deviation within the realm of normal probabilistic fluctuation. However, multiplying by 25,000 yields 27.1M, 28.3M, 27.8M, and 27.9M trials in each of the four settings configurations. If we pretend that the data for \cite{christensen:2013} came from an experiment where the detector settings were randomized prior to each trial, this four-way tally \emph{would} represent an improbably large deviation from an even four-way split for a 25\%-25\%-25\%-25\% random variable sampled 112.1M times.}, and noting that for both distributions, the sum of the probabilities $p($+0$ab')$, $p($0+$a'b)$, and $p($++$a'b')$ is exceeded by $p($++$ab)$.

\begin{table}[h]\caption{Empirical distributions for the experiments \cite{giustina:2013} (left) and \cite{christensen:2013} (right).}\label{t:empiricaldistributions}
\begin{tabular}{ r|c|c|c| }
 \multicolumn{1}{r}{}
  &  \multicolumn{1}{c}{++}
 &  \multicolumn{1}{c}{+0}
 &  \multicolumn{1}{c}{0+}  \\
 \cline{2-4}
 $ab$ & .050 & .021 & .029 \\
 \cline{2-4}
 $ab'$ & .054 & .017 & .157 \\
 \cline{2-4}
 $a'b$ & .056 & .165 & .023 \\
 \cline{2-4}
 $a'b'$  & .003 & .217 & .207 \\
 \cline{2-4}
 \end{tabular}
\hspace{1 cm}
\begin{tabular}{ r|c|c|c| }
 \multicolumn{1}{r}{}
  &  \multicolumn{1}{c}{++}
 &  \multicolumn{1}{c}{+0}
 &  \multicolumn{1}{c}{0+}  \\
 \cline{2-4}
 $ab$ & .044 & .026 & .026 \\
 \cline{2-4}
 $ab'$ & .049 & .020 & .162 \\
 \cline{2-4}
 $a'b$ & .051 & .172 & .019 \\
 \cline{2-4}
 $a'b'$  & .003 & .219 & .209 \\
 \cline{2-4}
 \end{tabular}
\end{table}

To measure the extremity of $J$, we conceptualize a CH experiment as a statistical hypothesis test, where we have two competing hypotheses:
\begin{eqnarray}\label{e:hypotheses}
H_0 &:&C_i \text{ has a local probability distribution for all } i \notag \\
H_A &:& \text{The } C_i \text{ follow quantum mechanics}.
\end{eqnarray}
Of interest is the probability of seeing a value of $J$ as or more extreme than what is observed, if we assume the null hypothesis $H_0$. This probability is the p-value. 

As the distribution of the $C_i$ variables can vary from trial to trial, the behavior of $J_k$ may change over time. However, under a local hidden variable theory, expression (\ref{e:use3}) tells us that $P(J_k = +1)\le {1\over 2}$ must always hold, no matter what has happened in the earlier instances of the $J_j$ variables for $j<k$. Intuitively, the best strategy for racking up ``+1" counts in a binary-output supermartingale sequence like $\{J_k\}$ would be to saturate the inequality (\ref{e:use3}) for each $k$; this point is used in \cite{barrett:2002} and shown to be true in \cite{bierhorst1:2013}. So if we take ${\mathcal L}$ to represent the set of local hidden variable theories falling under $H_0$, we have:
\begin{multline}\label{e:giustinap}
\sup_{ l \in \mathcal L} P(J\ge J_{obs} | l) =\\
 P(J\ge J_{obs} |  J_k \text{ i.i.d. with }P(J=+1)=1/2).
\end{multline}
The left side of (\ref{e:giustinap}) defines the p-value of an experiment resulting in an observed $J$ value of $J_{obs}$. The right side of (\ref{e:giustinap}) can be calculated exactly using the binomial distribution or approximated with the normal distribution for large $m$. 

As an example, we can apply this method to the raw data of \cite{giustina:2013} (with the caveat that this experiment did not meet all of the necessary experimental conditions laid out in the previous section). In \cite{giustina:2013}, $J$ is equal to $126,715$ and $m$ is $2,011,897$. These $J$ and $m$ values would represent a departure of 89 standard deviations from the local-hidden-variable expected value of $E(J)= 0$, if one appeals to the asymptotic normal distribution to estimate the exact binomial distribution. This figure is actually an improvement over the claimed 70 standard deviations in \cite{giustina:2013}.

Some of the justifications used to get to this point are complicated, but it is important to not lose sight of the following: the resulting rigorous statistical analysis is very straightforward to implement. One need only tally the trials that result in $\text{++}ab$, compute a second tally of trials that result in $\text{+0}ab'$, $\text{0+}a'b$, or $\text{++}a'b'$, and then decide whether the two numbers could have reasonably be generated by a source that was not biased towards producing $\text{++}ab$ tallies. The validity of this simple analysis is an important result of this paper.

\subsection{Clauser-Horne-Style Statistics}

Eberhard's $J$ is not the only statistic that can be used. Other statistics exploit other parts of the probability table that witness the deviation from local hidden variable theories. For instance, consider the following equations, which must hold for any local distribution for $C_i$ by the no-signaling principle and the equiprobable setting distribution:
\begin{eqnarray}\label{e:newNS1} 
p(\text{++}a b) + p(\text{+0} ab) &=& p(\text{++}ab') + p(\text{+0}ab')  \\
p(\text{++} a b) + p(\text{0+} a b) &=& p(\text{++} a' b) + p(\text{0+} a' b). \label{e:newNS3} 
\end{eqnarray}
(\ref{e:newNS1}) and (\ref{e:newNS3}) can be combined with (\ref{e:eberhardo}) to obtain, respectively,
\begin{flalign}\label{e:E2}
p(\text{++}ab') - p(\text{+0} ab)-p(\text{0+}a'b) -p(\text{++}a'b') \le 0\\
p(\text{++} a' b) - p(\text{+0}ab')- p(\text{0+} a b) -p(\text{++}a'b') \le 0, \label{e:E3}
\end{flalign}
which must be obeyed by local hidden variable theories. Either (\ref{e:E2}) or (\ref{e:E3}) can be substituted for (\ref{e:eberhardo}) as a starting point for the definition of $J_k$ in (\ref{e:Jdef}), yielding similarly useful $J$-style test statistics. One can also derive the original CH 74 inequality from these expressions: the sum of (\ref{e:E2}) and (\ref{e:E3}) is 
\begin{multline}\label{e:CH2}
p(\text{++}ab') + p(\text{++}a'b)-2p(\text{++}a'b')-p(\text{+0}ab) \\
-p(\text{+0}ab') -p(\text{0+}ab)- p(\text{0+}a'b)\le 0.
\end{multline}
In an experiment where $p(a)=p(b)={1\over 2}$, (\ref{e:CH2}) is equivalent to the original CH 74 inequality,
\begin{multline}\label{e:CH74}
p(+_1+_2|ab) + p(+_1+_2|ab')+p(+_1+_2|a'b) \\
- p(+_1+_2|a'b')-p(+_1|a)-p(+_2|b)\le 0.
\end{multline}
To see the equivalence, one can add and subtract the quantity $p(\text{++}ab')+p(\text{++}ab')+2p(\text{++}ab)$ from the left side of (\ref{e:CH2}) and then use the fact that 
\begin{eqnarray*}
p(+_1a)&=&p(\text{++}ab)+p(\text{+0}ab)+p(\text{++}ab')+p(\text{+0}ab')\\
p(+_2b)&=&p(\text{++}ab)+p(\text{0+}ab)+p(\text{++}a'b)+p(\text{0+}a'b).
\end{eqnarray*}
Hence if one were to derive a statistic based directly on equation (\ref{e:CH2}), this could be interpreted as a direct measure of violation of the original CH 74 inequality.

Similar methods can be used to generate many other statistics, and a classification of all the relevant statistics is given in Appendix \ref{s:classification}. However, it turns out that some of these statistics require a more complicated analysis than $J$, and one example is the statistic generated by (\ref{e:CH2}).

To define a statistic based on (\ref{e:CH2}), let $M$ be the subset of outcomes that appear in (\ref{e:CH2}), and if we re-define $T_k$ to be the $k$th $C_i$ in $M$, we consider the following ``step" variable
\begin{equation*}
Ch_k = \begin{cases} +1, & \text{if $T_k=$++$ab'$ or ++$a'b$} \\ 
 -1, & \text{if $T_k=$+0$ab$, 0+$ab$, +0$ab'$, or 0+$a'b$ }\\
 -2, & \text{if $T_k=$++$a'b'$}. \end{cases}
\end{equation*}
For bookkeeping purposes, we can define $Ch_k$ to be $-2$ if fewer than $k$ trials assume a value in the set $M$. Define the new test statistic $Ch$ as the sum $\sum_{k=1}^m Ch_k$, where $m$ is the total number of $C_i$ variables that take a value in the set $M$ over the course of the experiment. Now, similarly to our work with $J_k$, it can be shown that
\begin{multline}\label{e:use4}
P(Ch_k=+1|\mathcal{CH}_k)\le\\
 P(Ch_k=-1|\mathcal{CH}_k) + 2P(Ch_k=-2|\mathcal{CH}_k),
\end{multline}
where $\mathcal{CH}_k$ is any positive-probability event consisting of outcomes of $Ch_i$ variables for $i<k$.

In contrast to our analysis of $J$, there are now many different local distributions that saturate the inequality (\ref{e:use4}), with possibilities including
\begin{eqnarray}\label{e:saturate1}
P(Ch_k=+1) = {1\over 2} \text{ and } P(Ch_k=-1) = {1\over 2},  \\
P(Ch_k=+1) = {2\over 3} \text{ and } P(Ch_k=-2) = {1\over 3},  \label{e:saturate2}
\end{eqnarray}
as well as any convex combination of (\ref{e:saturate1}) and (\ref{e:saturate2}). This is a consequence of working with a supermartingale sequence with output in a non-binary set. The optimal local strategy for generating a large $Ch$ may involve switching between these two distributions for different trials, and so the analysis is more complicated than the analysis of $J$. (Note that a local hidden variable theory can achieve distribution (\ref{e:saturate1}) by, for instance, repeating the deterministic strategy $v_9$ from Table \ref{t:deterministicstrategies} in Appendix \ref{s:classification}, and can achieve distribution (\ref{e:saturate2}) by repeating the deterministic strategy $v_{1}$.)

One way to proceed is to follow a strategy first outlined in \cite{GILL}. This work demonstrated that supermartingales naturally arise when analyzing Bell test experiments with possible memory effects, and suggested the use of the Azuma-Hoeffding inequality \cite{HOEFFDING,AZUMA} to bound the probability that these martingale-based statistics exceed certain values. As noted in \cite{zhang:2013}, the tightest relevant Azuma-Hoeffding-style bound can be obtained from Theorem 6.1 in  \cite{mcdiarmid:1989}, which in our situation yields 
\begin{equation}\label{e:mcdiarmid}
P(Ch\ge mt) \le \left[ {2\overwithdelims () 2+ t}^{2+t\over 3} {1\overwithdelims () 1-t}^{1-t\over 3}\right]^m, 
\end{equation}
where $m$ is the number of $Ch_k$ steps and $t$ is chosen so that $mt$ is the desired cut-point. The above constraint can be used to upper-bound p-values for the extremity of $Ch$.

If we apply this method to the raw data of \cite{christensen:2013}, $Ch$ would be $4,258$ after $m=131,116$ movements of the test statistic. Applying the inequality (\ref{e:mcdiarmid}), one obtains $P(Ch \ge 4,258) \le 8.0\times 10^{-16}$, a figure exceeding the significance level claimed by the authors of \cite{christensen:2013}. However, care should be taken as the distribution over measurement settings in \cite{christensen:2013} exhibit a slight deviation from equiprobability (see [18]). A generalization of the analysis to non-equiprobable setting distributions is given in Appendix \ref{s:mod}.

\subsection{Calculating an Exact P-value for Non-Binary Supermartingales}\label{s:exact}

For the values of $Ch$ and $m$ in the previous paragraph, (\ref{e:mcdiarmid}) easily suffices to demonstrate statistical significance. However, this bound is not tight (even asymptotically), and for different values of $Ch$ and $m$ it may be desirable to calculate sharper bounds on this probability. This will be especially useful for the design of an experiment in which only a limited number of trials is possible. The exact probability can be calculated by a new method described here.

$Ch$ is a random walk on the integers starting at zero, with the $k$th step given by $Ch_k$, resulting in a move either up 1, down 1, or down 2. The goal for the local hidden variable theory is to finish at or above a specified cut-point $L$ after $m$ steps (in \cite{christensen:2013}, $L=4,258$ and $m=131,116$), and the p-value is the probability of attaining this goal. The trick for determining the best strategy for finishing at or above $L$ is to trace backwards from the end.

\begin{table}[h]\caption{Optimal Strategies and Probabilities of Eventual Success at Various Locations, $Ch$ Random Walk}\label{t:backtrace}
\begin{tabular}{r||cc|cc|cc|}
\multicolumn{1}{c||}{Location} & \multicolumn{2}{c | }{$m-2$ step}   & \multicolumn{2}{c | }{$m-1$ step}   & \multicolumn{2}{c |}{m step}\\
 \multicolumn{1}{c||}{of $Ch$} & O.S. & $P(S)$ & O.S. & $P(S)$ & O.S. & $P(S)$\\
\hline
 \multicolumn{1}{c||}{$L+3$} & any & 1 & any & 1 & - & 1\\
  \multicolumn{1}{c||}{$L+2$} & (\ref{e:saturate1}) & 1 & any & 1 & - & 1\\
   \multicolumn{1}{c||}{$L+1$} &  (\ref{e:saturate2}) & ${8\over 9}$  & (\ref{e:saturate1})      & 1 & - & 1\\
 \multicolumn{1}{c||}{$L$}     & (\ref{e:saturate1}) & ${5\over 6}$      & (\ref{e:saturate2})        & ${2\over 3}$ & - & 1\\
 \multicolumn{1}{c||}{$L-1$} & (\ref{e:saturate2}) & ${4\over 9}$      & (\ref{e:saturate2})        & ${2\over 3}$ & - & 0\\
 \multicolumn{1}{c||}{$L-2$} & (\ref{e:saturate2}) & ${4\over 9}$    & any   & 0  & - & 0 \\
 \multicolumn{1}{c||}{$L-3$} & any & 0      & any    & 0 & - & 0  \\
\hline
\end{tabular}
\end{table}

The first few iterations of this process are displayed in Table \ref{t:backtrace}. At the final, $m$th step, the random walk $Ch$ has ``succeeded" if it is at or above $L$, and ``failed" otherwise. Hence the probability of eventual success is either 1 or 0, respectively. 

Now, move back a step, to the $m-1$ step. For a given location of the random walk, we want to calculate the optimal probability distribution for $Ch_m$ that maximizes the probability of eventual success. Since $Ch_m$ must take a value from the set $\{-2, -1, +1\}$, the only interesting locations for the random walk at step $m-1$ are $L-1$, $L$, and $L+1$. Any higher location translates to certain eventual success, and any lower location translates to certain eventual failure.

If the probability distribution for $Ch_m$ is induced by a local hidden variable theory, it must satisfy (\ref{e:use4}). Only distributions saturating this inequality will be able to maximize the probability of eventual success. To see why, consider the random walk sitting at location $L-1$ at the $m-1$ step. Any distribution for $Ch_m$ over the set $\{-2, -1, +1\}$ that does not saturate the inequality can be improved upon by taking some of the probability away from $-1$ or $-2$ and adding it to $+1$; this clearly will increase the probability of eventual success. Importantly, this would still be true if we were to replace the 0's and 1's in the last column with any monotone-increasing collection of numbers. (This observation will be important for applying the argument to the earlier steps $m-2$, $m-3$, etc.)

Now, any distribution saturating the inequality (\ref{e:use4}) will be a convex combination of the two distributions (\ref{e:saturate1}) and (\ref{e:saturate2}). Hence, the optimal probability of eventual success can be obtained by either using (\ref{e:saturate1}) or (\ref{e:saturate2}), because any convex combination of these two distributions will have an eventual probability of success that is the weighted average of the probability of success using (\ref{e:saturate1}) and the probability of success using (\ref{e:saturate2}). As noted earlier, either one of the distributions (\ref{e:saturate1}) and (\ref{e:saturate2}) can be implemented by local hidden variable theories.

Since the optimal distribution for $Ch_m$ at any particular location is always either (\ref{e:saturate1}) or (\ref{e:saturate2}), the optimal strategies for all locations at step $m-1$ can be calculated effectively. For instance, we see the random walk that finds itself at location $L+1$ at the penultimate step should opt for the 50-50 strategy (\ref{e:saturate1}) that puts equal weight on moving up 1 and down 1, because this strategy ensures eventual success. On the other hand, the random walk that finds itself at location $L$ will want to select strategy (\ref{e:saturate2}), which gives a best-possible ${2\over 3}$ chance of ending at/above $L$. The entire $m-1$ column is thus filled. Importantly, the probabilities of eventual success will continue to be monotone in the $m-1$ column, which allows the process to be repeated for step $m-2$, then $m-3$, etc. The reason that the probabilities of eventual success for each successively-filled-in column continue to be monotone is this: if the probability for eventual success at location $x$ using distribution (\ref{e:saturate1}) is $p$, then the probability for success at $x+1$ using strategy (\ref{e:saturate1}) must be at least $p$, because of the monotonicity of the probabilities of success at the next step (the previously-filled-in column), and a similar thing can be said for the distribution (\ref{e:saturate2}).

A computer can quickly extrapolate back to the first step, and the probability of eventual success for the random walk at location $0$ on step $0$ will be the \emph{exact} p-value for the extremity of the $Ch$ statistic.

As can be seen just from the first few steps of the process in Table \ref{t:backtrace}, there is something to be gained by employing different strategies at different steps of the process -- i.e., exploiting the memory loophole. This is an interesting phenomenon, because this is not true for the binary variable $J_k$ (local hidden variable theories exploiting memory cannot beat an i.i.d. strategy for achieving large values of $J$). We can get a better sense for why this is by considering what a \emph{memoryless} local hidden variable theory would do to maximize $Ch$. If so restricted, it makes intuitive sense to employ distribution (\ref{e:saturate2}) for all trials, as this distribution has the largest variance, and is thus more likely to produce positive fluctuations exceeding a given cut-point. But if the hidden variable theory ``gets lucky" with a large positive fluctuation (above the quantum prediction) in an initial set of experimental trials, only a theory with memory could employ the following strategy: switch to the lower-variance distribution (\ref{e:saturate1}) to reduce the size of negative fluctuations, and thus lock in the gains for a longer period of time while waiting for the experiment to end.

These complicating effects can be avoided by working only with $J$-style binary statistics. However, this is only possible if the setting probabilities are kept at (or near) 50-50 at both ends. As is seen in Appendix \ref{s:mod}, other setting probability distributions necessitate the use of more complicated non-binary martingale techniques. Furthermore, in other Bell scenarios that involve different numbers of detectors, settings, and/or outcomes, non-binary martingale statistics may be unavoidable, and the procedure for finding exact p-values described here is widely applicable.

\subsection{Summary of Statistical Analysis}\label{s:statsummary}

Exact p-values can now be calculated for all of the test statistics discussed in this section, using the back-tracing method for $Ch$ and the binomial distribution for $J$ and related statistics $J_{(\ref{e:E2})}$ and $J_{(\ref{e:E3})}$ (which test (\ref{e:E2}) and (\ref{e:E3}), respectively, instead of (\ref{e:eberhardo})). To summarize these results and to compare the performance of these statistical tests, an experiment of $n=100,000$ non-00 trials was simulated according to empirical distributions in Table \ref{t:empiricaldistributions}, and the results for different statistical tests are summarized in Table \ref{t:simulate}.

\begin{table}[h]\caption{Statistical Outcomes for Simulated 100,000 Trial Experiment}\label{t:simulate}
\begin{tabular}{rcccccc}
& \multicolumn{3}{c}{Giustina \emph{et. al} \cite{giustina:2013}}   & \multicolumn{3}{c }{Christensen \emph{et. al} \cite{christensen:2013}}   \\
Statistic & Value & $m$ & p-value & Value & $m$ & p-value \\
\hline
\multicolumn{1}{|c|}{$J$} & $591$ & $9,380$ & \multicolumn{1}{c|}{$5.17\times10^{-10}$} & $206$ & $8,624$ & \multicolumn{1}{c|}{$.0136$}\\
\multicolumn{1}{|c|}{$J_{(\ref{e:E2})}$} & $573$ & $10,175$ & \multicolumn{1}{c|}{$7.06\times10^{-9}$} & $202$ & $9,696$ & \multicolumn{1}{c|}{$.0206$} \\
\multicolumn{1}{|c|}{$J_{(\ref{e:E3})}$} & $562$ & $10,545$ & \multicolumn{1}{c|}{$2.20\times10^{-8}$} & $245$ & $9,937$ & \multicolumn{1}{c|}{$.0072$}\\
\multicolumn{1}{|c|}{$Ch$} & $1,135$ & $20,395$ & \multicolumn{1}{c|}{ $9.90\times 10^{-9}$ } & $447$ & $19,359$ & \multicolumn{1}{c|}{$.0136$}\\
\hline
\end{tabular}
\end{table}

It should be mentioned that while the distribution for \cite{giustina:2013} produces stronger Bell violations in Table \ref{t:simulate}, the violation would probably be weakened if the data on which the empirical distribution was based had been filtered into discrete time intervals. A filtering step is unnecessary for the data of \cite{christensen:2013}, as the experimental design of \cite{christensen:2013} incorporated discrete trials.

Table \ref{t:simulate} reveals a few interesting things. First, we note that the new method allowing exact calculations of the $Ch$ p-values represent a meaningful improvement over the un-sharp upper bound given by the inequality (\ref{e:mcdiarmid}): for the exact quantity $9.90\times 10^{-9}$ in the table above, the bound given by (\ref{e:mcdiarmid}) is $1.19 \times 10^{-7}$; for the quantity .0136, this bound is .0750.

Another interesting feature is that the choice of statistic can make a difference in the p-value. To illustrate how this comes about, consider the following hypothetical distribution, which is non-signaling and non-local: 

\begin{table}[h]
\begin{tabular}{ r|c|c|c| }
 \multicolumn{1}{r}{}
  &  \multicolumn{1}{c}{++}
 &  \multicolumn{1}{c}{+0}
 &  \multicolumn{1}{c}{0+}  \\
 \cline{2-4}
 $ab$ & .110 & .002 & {\tiny 0} \\
 \cline{2-4}
 $ab'$ & .012 & .100 & {\tiny 0} \\
 \cline{2-4}
 $a'b$ & {\tiny .110} & {\tiny .272} & 0 \\
 \cline{2-4}
 $a'b'$  & 0 & {\tiny .382} & {\tiny .012} \\
 \cline{2-4}
 \end{tabular}
\end{table}

\noindent The probabilities that appear in inequalities (\ref{e:eberhardo}) and (\ref{e:E2}) have been emphasized. If an experiment is run 10,000 times and the outcomes roughly coincide with the probabilities given by the above table, then the statistics $J$ and $J_{(\ref{e:E2})}$ should both take values close to 100. However, the p-values will be very different, because $J=\sum_{k=1}^m J_k$ will be obtained by adding up 1,100 ``$+1$" outcomes and subtracting 1,000 ``$-1$" outcomes (so $m=2,100$), while $J_{(\ref{e:E2})}$ is obtained by adding up 120 ``$+1$" outcomes and subtracting 20 ``$-1$" outcomes (so $m_{(\ref{e:E2})}=140$). It is of course very unlikely to get a net imbalance of 100 over the course of 140 trials sampled from a 50-50 distribution (the best a local hidden variable theory can do), but far less unlikely to get the same absolute imbalance over the course of 2,100 trials.

This example demonstrates that a useful statistic is one that accumulates a large positive value with as little extraneous up-and-down movement as possible. When planning an experiment, one can use this idea to choose which statistic to employ by studying the quantum prediction for the experimental distribution.

\section{Conclusion}\label{s:conclusion}

We have demonstrated how to perform a correct statistical analysis of a loophole-free Clauser-Horne experiment without making any tacit assumptions. Specifically, we have shown how to extract data from the subset of relevant experimental trials and statistically analyze it while not ignoring the possibility of memory effects. These methods have been applied to the raw data of recent Clauser-Horne experiments \cite{giustina:2013,christensen:2013} and shown to perform well.

We have also introduced a new method for calculating exact p-values for any supermartingale statistic that takes an output in a finite set. This improves on the best-previously-known Azuma-Hoeffding bounds, and the method can be applied to other scenarios in which supermartingale probabilities may be of interest. This includes other Bell scenarios and applications such as device-independent random number expansion \cite{pironio:2010} and key distribution \cite{BHK} when memory effects are taken in to account.

The experimental framework and statistical analysis described in this paper are sufficient for a loophole-free test of Bell's inequality. While a local hidden variable theory could have exploited the locality loophole to generate the data of \cite{giustina:2013,christensen:2013}, a future experiment realizing the framework put forth in Section \ref{s:basics} of this paper and achieving statistically significant p-values would eliminate any remaining possibility for a local hidden variable theory of photons.



\begin{acknowledgments}
The author would like to thank Mike Mislove for many helpful discussions of the material presented here.
\end{acknowledgments}

\bibliography{c:/Users/pbierhor/Dropbox/metabib}

\appendix

\vspace{.5 cm}

\begin{center}
{\bf Appendix}
\end{center}

\vspace{-.8cm}

\subsection{Block-Measurement Designs and Spacelike Separation}\label{s:loopholes}

Both experiments \cite{giustina:2013} and \cite{christensen:2013} did not enforce spacelike separation between detection events. Additionally, these experiments measured multiple successive trials in the same measurement configuration. For instance, \cite{christensen:2013} operated the experiment in a fixed measurement configuration for about 25,000 experimental trials (in 1 second), then randomly re-assigned the measurement settings and measured for another 25,000 trials, repeating this process 4450 times. While it may theoretically be possible to implement a loophole-free test of a Bell inequality with such a ``block-measurement" experimental design, this scenario poses additional challenges for statistical analysis. These challenges are more fully explored in \cite{bierhorst2:2013}, but we touch on the main points here.

The main problem with a block-measurement design is the added difficulty of ruling out the locality loophole. If one were to take an experiment like \cite{christensen:2013} and separate the two detectors so that simultaneous trials were fully space-like separated from each other, that would require a separation of about 12 km between the two detectors, based on the duration of the trials (.04 ms) and the speed of light. This is within the realm of experimental feasibility. However, if the detector settings are not re-randomized prior to each trial, a local hidden variable could generate just about any desired distribution by employing the following strategy: at the beginning of a block of measurements, generate a few throwaway outcomes at detector 1, waiting for information about the setting at detector 2 to arrive via sub-luminal signals. Once that information arrives, the hidden variable has full information about all the settings, and can simulate any desired distribution for the rest of the block.

One possibility for addressing this is to increase the degree of separation so that final measurement event at the end of a block takes place before a not-faster-than-light signal could have arrived from the very first measurement event of the same block at the other end of the experiment. With a block duration of one second, this would call for a technically-infeasible separation distance on the order of Earth's distance from the Moon. Therefore, any feasible loophole-free Bell test should focus instead on re-randomizing the measurement settings prior to each trial window.

Furthermore, even if this degree of separation could be achieved, it is an open question whether there is a statistical analysis that could distinguish between quantum mechanics and local theories in a block-measurement design. A proposed solution in the online supporting material of \cite{christensen:2013} has flaws, as detailed in \cite{bierhorst2:2013}, but an upcoming work may soon resolve this question \cite{knill:2014}. 

\subsection{Derived Supermartingales}\label{s:subsequence}

Here we prove a result in Section \ref{s:classification} that allows for the derivation of expression (\ref{e:use3}). This is the result that allows us to disregard trials of the experiment that do not appear in the chosen Bell inequality. The proof is given in general form so that it can apply not just to $J$ but also to $Ch$ or any other similarly-conceived statistic. Hence the proposition can also be used to justify expression (\ref{e:use4}).

To apply the proposition below explicitly to the $J$ statistic, one would have $X_i=J'_i$, $U=\{-1,0,+1\}$, $W=\{-1,+1\}$, expression (\ref{e:mathXconstraint}) corresponding to (\ref{e:use2}), $\mathcal X=\mathcal {J}'$, $Y_k=J_k$, $d=-1$, expression (\ref{e:mathYconstraint}) corresponding to (\ref{e:use3}), and $\mathcal Y=\mathcal {J}$.

\medskip

\noindent\emph{Proposition:} Let $\{X_i\}_{i=1}^\infty$ be a sequence of random variables taking values in a finite set $U$. Suppose the $X_i$ satisfy the following condition: there is a collection of real constants $\{c_1, ..., c_n\}$ and a subset $W=\{s_1, ..., s_n\}\subseteq U$ such that for any fixed $i$, 
\begin{equation}\label{e:mathXconstraint}
\sum_{j=1}^n c_j P(X_i=s_j\mid \mathcal X) \le 0,
\end{equation}
where $\mathcal X$ is allowed to be any fixed event of the form $(X_1, ..., X_{i-1})=\vec u$, $\vec u \in U^{i-1}$ for which $P(\mathcal X)>0$.

Now suppose we define the new random variable sequence $\{Y_k\}_{k=1}^\infty$ as follows:
\begin{equation}\label{e:useA1}
Y_k = \text{ the value of the }k\text{th }X_i \text{ taking a value in the set $W$}
\end{equation}
and we define $Y_k$ to be $d$, where $d$ is a particular choice of the $s_j$ values, if fewer than $k$ of the $X_i$ take a value in the set $W$. 

Then if the $c_j$ corresponding to $s_j=d$ in (\ref{e:mathXconstraint}) is not positive, the following holds for any fixed value of $k$:
\begin{equation}\label{e:mathYconstraint}
\sum_{j=1}^n c_j P(Y_k=s_j\mid \mathcal Y) \le 0,
\end{equation}
where $\mathcal Y$ is any fixed event of the form $(Y_1, ..., Y_{k-1})=\vec w$, $\vec w \in W^{k-1}$ for which $P(\mathcal Y)>0$.

\medskip

\noindent\emph{Proof:} The event $\mathcal Y$ can be broken up into a disjoint union of two events,
\begin{equation*}
\mathcal Y = \mathcal Y^0 \cup \mathcal Y',
\end{equation*}
where $\mathcal Y^0$ consists of the event $(Y_1, ..., Y_{k-1})=\vec w$ intersected with the event ``fewer than $k$ of the $X_i$ variables assume a value in the set $W$," and $\mathcal Y'$ consists of the event $(Y_1, ..., Y_{k-1})=\vec w$  intersected with the event ``at least $k$ of the $X_i$ variables assume a value in the set $W$." Note that it is not necessary for both $\mathcal Y^0$ and $\mathcal Y'$ to be positive-probability events. So, we can write 
\begin{multline}\label{e:useA2}
\sum_{j=1}^n c_j P(Y_k=s_j \cap \mathcal Y) =\\
 \sum_{j=1}^n c_j P(Y_k=s_j \cap \mathcal Y^0) + \sum_{j=1}^n c_j P(Y_k=s_j \cap \mathcal Y') .
\end{multline}
Note that we have temporarily moved to a discussion joint probabilities with ``$\cap$" instead of the conditional probabilities appearing on the left side of (\ref{e:mathYconstraint}). We will return to conditional probabilities in the last step of the proof.

The first sum on the right side of equation (\ref{e:useA2}) will be nonpositive, because if the event $\mathcal Y^0$ occurs, $Y_k$ necessarily equals $d$. Then, the only potentially-positive probability in the sum will be $P(Y_k=d \cap \mathcal Y^0)$, and by assumption the coefficient of this term is nonpositive.

Now we move to the second sum on the right side of equation (\ref{e:useA2}). Our strategy is to break up the event $\mathcal Y'$ into a disjoint union of events that refer only to the $X_i$ variables, and then use (\ref{e:mathXconstraint}) to bound the terms. We will break up $\mathcal Y'$ into constituent events as follows. Let $I$ be an indexing set, and let $\{\mathcal X_i^{l}\}_{i\in I, l\in(1,...,n)}$ be a collection of sets of the form $\mathcal X_i^{l} = \{(X_1,..., X_{m_i}) = \vec u_i^l\}$, where $m_i$ is an integer greater than or equal to $k$ ($m_i$ may vary for different choices of $i\in I$), and $\vec u_i^l$ is an $m_i$-dimensional vector satisfying the following conditions: the first $m_i-1$ entries are uniquely fixed by the choice of $i\in I$, exactly $k$ entries of $\vec u_i^l$ assume a value in the set $W$ with the first $k-1$ of these comprising the vector $\vec w$, and the final entry of $\vec u_i^l$ is $s_l$. Let $I$ index precisely the choices of $i$ for which $\cup_{l=1}^n\mathcal X_i^l$ intersected with $\mathcal Y'$ is a set of positive probability. (If $I$ is empty, we can immediately skip ahead to (\ref{e:useA6}), so for the following argument we assume that $I$ is nonempty.) Now we can write $\mathcal Y' = \cup_{i\in I}\cup_{l=1}^n \mathcal X_i^l$, where the union is disjoint. We see that
\begin{eqnarray}\label{e:useA4}
\sum_{j=1}^n c_j P(Y_k=s_j\cap \mathcal Y') &=& \sum_{j=1}^n c_j \sum_{i\in I}\sum_{l=1}^n P(Y_k=s_j\cap \mathcal X_i^l)\notag\\
&=& \sum_{j=1}^n c_j \sum_{i\in I} P(Y_k=s_j\cap \mathcal X_i^j)\notag\\
&=& \sum_{j=1}^n c_j \sum_{i\in I} P(\mathcal X_i^j),
\end{eqnarray}
because $P(Y_k=s_j\cap \mathcal X_i^k)=0$ unless $j=l$, and the event $\{Y_k=s_j\}$ contains the event $\mathcal X_i^j$. Now, if we re-write (\ref{e:useA4}) as 
\begin{equation*}
\sum_{j=1}^n  \sum_{i\in I} c_j P(\mathcal X_i^j),
\end{equation*}
the order of summation can be switched because the sum converges absolutely, as we are working with a collection of disjoint events in a probability space, so this yields
\begin{equation}\label{e:useA5}
 \sum_{i\in I} \sum_{j=1}^n  c_j P(\mathcal X_i^j).
\end{equation}
Now, we can break up each $\mathcal X_i^j$ into the intersection of two events $\{X_{m_i}=s_j\}\cap  \{(X_1,..., X_{m_i-1}) = \vec u_i^{*}\}$, where $\vec u_i^{*}$ is obtained from $\vec u_i^{j}$ by truncating the last entry (recall that the first $m_i-1$ entries of $\vec u_i^j$ are the same for any choice of $j$, so $\vec u_i^{*}$ does not depend on $j$). We see that $P\left((X_1,..., X_{m_i-1}) = \vec u_i^{*}\right)>0$ by the assumption that for any choice of $i$, $\cup_{j=1}^n\mathcal X_i^j$ intersected with $\mathcal Y'$ is a set of positive probability. Now, (\ref{e:mathXconstraint}) implies that 
\begin{equation*}
\sum_{j=1}^n c_j P(\{X_{m_i}=s_j\}\mid  (X_1,..., X_{m_i-1}) = \vec u_i^{*})\le 0,
\end{equation*}
or, multiplying both sides of the inequality by $P\left((X_1,..., X_{m_i-1}) = \vec u_i^{*}\right)$,
\begin{equation*}
\sum_{j=1}^n c_j P(\{X_{m_i}=s_j\}\cap  (X_1,..., X_{m_i-1}) = \vec u_i^{*})\le 0.
\end{equation*}
The above expression tells us that (\ref{e:useA5}) is bounded above by zero. Recalling that (\ref{e:useA5}) is equivalent to the second sum in (\ref{e:useA2}), we can now say that
\begin{equation}\label{e:useA6}
\sum_{j=1}^n c_j P(Y_k=s_j \cap \mathcal Y)\le 0,
\end{equation}
and dividing both sides of the above expression by $P(\mathcal Y)$ (which by assumption is nonzero), this is equivalent to 
\begin{eqnarray*}
\sum_{j=1}^n c_j P(Y_k=s_j \mid \mathcal Y)\le 0.\quad  \quad \hfill \Box
\end{eqnarray*}

\subsection{Classification of All Relevant Constraints}\label{s:classification}

Here, we categorize all possible CH-type Bell inequalities that are relevant to the scenario of the paper: $2\times 2\times 2$ experiments in which measurement settings are equiprobable and only the non-00 outcomes are considered for statistical purposes.

Each trial of a CH experiment results in one of 16 outcomes, but only a subset of the trials will result in one of the 12 non-00 outcomes. If we denote the set of these 12 outcomes as $K$, we can define a new sequence of random variables $S_j$ as follows: 
\begin{eqnarray*}
C_i &=& \text{ the outcome of the $i$th trial}\\
S_j &=& \text{ the $j$th $C_i$ that takes a value in $K$.}
\end{eqnarray*}
The last column of Table \ref{t:trialsubset} illustrates how the values of $S_j$ are set. Constraints on the $S_j$ variables will be the kind of constraints that can be adapted to form random variable sequences like $\{J_k\}$, where the goal is to not depend on the occurrence of 00 outcomes. It is not assumed that the $C_i$ form independent and/or identically distributed sequences, so possible memory effects are not ruled out as we calculate the class of acceptable distributions for the $S_j$ random variables under local hidden variable theories.

It was shown by Fine \cite{fine:1982} that for a $2\times2\times2$ experiment, any local probability distribution for a $C_i$ random variable can be modeled by a convex combination of the 16 deterministic strategies, where a ``deterministic strategy" is an assignment of the two outcomes $\{+,0\}$ to each of the four settings $\{a, a', b, b'\}$. We denote the set of deterministic strategies as $\{v_k\}_{k=1}^{16}$, which are listed in Table \ref{t:deterministicstrategies}.

\begin{widetext}
\begin{center}
\begin{table}[h]\caption{Deterministic Strategies}\label{t:deterministicstrategies}\small
\begin{tabular}{ rcccc||c|c|c|c||c|c|c|c||c|c|c|c||c|c|c|c|| }
& & & & & \multicolumn{4}{c||}{$ab$} & \multicolumn{4}{c||}{$ab'$} & \multicolumn{4}{c||}{$a'b$} & \multicolumn{4}{c||}{$a'b'$}\\
&$a$ & $a'$ & $b$ & $b'$ & ++ & +0 & 0+ & 00 &    ++ & +0 & 0+ & 00     & ++ & +0 & 0+ & 00 &         ++ & +0 & 0+ & 00 \\ 
\cline{2-21}
\multicolumn{1}{r|}{$v_1$} & + & + & + & +    & X &        &      &                &      X &       &      &                &   X &       &        &                      & X    &       &      &    \\
 \multicolumn{1}{r|}{$v_2$} & 0 & + & + & +     &   &        &    X  &                &       &       &   X   &               &   X &       &        &                       & X   &       &      &       \\
 \multicolumn{1}{r|}{$v_3$} & + & 0 & + & +    &X   &        &      &                &   X   &       &      &               &     &       &     X  &                       &     &       &    X  &       \\
 \multicolumn{1}{r|}{$v_4$} & + & + & 0 & +      &   &   X     &     &                &  X     &       &       &               &     &    X   &        &                       &  X  &       &      &       \\
 \multicolumn{1}{r|}{$v_5$} & + & + & + & 0      &X  &        &     &                &       &    X   &       &               &  X   &       &        &                       &    &  X     &      &       \\
 \multicolumn{1}{r|}{$v_6$} & 0 & 0 & + & +     &   &        &  X   &                &       &       &    X   &               &     &       &      X  &                       &    &       &   X   &       \\
 \multicolumn{1}{r|}{$v_7$} & + & + & 0 & 0 &   & X       &     &                &       &    X   &       &               &     &  X     &        &                       &    &    X   &      &       \\ 
 \multicolumn{1}{r|}{$v_8$} & 0 & + & + & 0     &   &        & X    &                &       &       &       &  X             & X    &       &        &                       &    &    X   &      &       \\
 \multicolumn{1}{r|}{$v_9$} & + & 0 & 0 & +      &   &  X      &     &                & X      &       &       &               &     &       &        &  X                     &    &       &   X   &       \\
 \multicolumn{1}{r|}{$v_{10}$} & + & 0 & + & 0 & X  &        &     &                &       &    X   &       &               &     &       &   X     &                       &    &       &      &    X   \\
 \multicolumn{1}{r|}{$v_{11}$} & 0 & + & 0 & +     &   &        &     &    X            &       &       &  X     &               &     &  X     &        &                       &  X  &       &      &       \\    
 \multicolumn{1}{r|}{$v_{12}$} & 0 & 0 & 0 & + &   &        &     & X               &       &       &    X   &               &     &       &        & X                      &    &       &   X   &       \\
 \multicolumn{1}{r|}{$v_{13}$} & 0 & 0 & + & 0 &   &        &   X  &                &       &       &       & X              &     &       &   X     &                       &    &       &      &   X    \\
 \multicolumn{1}{r|}{$v_{14}$} & 0 & + & 0 & 0 &   &        &     &  X              &       &       &       &  X             &     &    X   &        &                       &    &  X     &      &       \\
 \multicolumn{1}{r|}{$v_{15}$} & + & 0 & 0 & 0  &   &   X     &     &                &       &  X     &       &               &     &       &        &  X                     &    &       &      &   X    \\           
 \multicolumn{1}{r|}{$v_{16}$} & 0 & 0 & 0 & 0 &   &        &     &   X             &       &       &       &  X             &     &       &        &  X                     &    &       &      &  X     \\    \cline{2-21}
 \end{tabular}
\end{table}
\end{center}
\end{widetext}

To understand this table, examine the state labeled $v_{11}$ as an example. This local hidden variable sends out the state, or ``strategy," $(a, a', b, b') = (0, +, 0, +)$. This is essentially an instruction set: at detector 1, yield ``0" if the setting is $a$ and trigger ``+" if the setting is $a'$; at detector 2, yield ``0" if the setting is $b$ and trigger ``+" if the setting is $b'$. Hence if the setting configuration is $ab$, the result is $00$, whereas $ab'$, $a'b$, or $a'b'$ would yield results $0$+, +0, or 00 respectively, as shown in the table. If the setting probability distributions are both 50-50 and independent (as is assumed), the distribution for the $C_i$ random variable can be determined, and it is given on the left side of Table \ref{t:strategies}.

Now, if \emph{all} the $C_i$ are governed by the particular state $v_{11}$, one can deduce that the $S_j$ will follow the probability distribution given on the right side of Table \ref{t:strategies}. The probability distributions for $C_i$ and $S_j$ induced this way by a $v_{k}$ state are denoted $\vec{v_{k}}$ and $\vec{v_{k}}^{**}$, respectively.

\begin{table}[h]\caption{Probability distributions $\vec{v_{11}}$ and $\vec {v_{11}}^{**}$ induced by deterministic strategy $v_{11}$=(0, +, 0, +) for the random variables $C_i$ (left) and $S_j$ (right)}\label{t:strategies}
\begin{tabular}{ r|c|c|c|c| }
 \multicolumn{1}{r}{}
  &  \multicolumn{1}{c}{++}
 &  \multicolumn{1}{c}{+0}
 &  \multicolumn{1}{c}{0+}
  & \multicolumn{1}{c}{00} \\
 \cline{2-5}
 $ab$ & 0  & 0 & 0 & 1/4 \\
 \cline{2-5}
 $ab'$ &  0 & 0 & 1/4 & 0 \\
 \cline{2-5}
 $a'b$ & 0 &  1/4 & 0 & 0 \\
 \cline{2-5}
 $a'b'$ &  1/4 & 0 & 0 & 0 \\
 \cline{2-5}
 \end{tabular}
\hspace{1 cm}
\begin{tabular}{ r|c|c|c| }
 \multicolumn{1}{r}{}
  &  \multicolumn{1}{c}{++}
 &  \multicolumn{1}{c}{+0}
 &  \multicolumn{1}{c}{0+}  \\
 \cline{2-4}
 $ab$ & 0 & 0 & 0 \\
 \cline{2-4}
 $ab'$ & 0 & 0 & 1/3 \\
 \cline{2-4}
 $a'b$ & 0 & 1/3 & 0 \\
 \cline{2-4}
 $a'b'$  & 1/3 & 0 & 0 \\
 \cline{2-4}
 \end{tabular}
\end{table}

Instead of choosing to send just one particular strategy $v_k$, a local hidden variable theory could alternatively choose to send a mixture of various strategies according to a probability distribution. Then the distribution of a particular $C_i$ would be a convex combination of the 16 $v_k$-induced distributions $\vec{v_k}$. Fine's result \cite{fine:1982} implies that all local probability distributions for $C_i$ can be realized by such a convex combination. We want to lift this result to the variables of interest, the $S_j$ variables.

Before proceeding, note that there is one particular $v_k$ state that does not induce a distribution for $S_j$: the particular strategy $v_{16}=(0,0,0,0)$. If repeated, $v_{16}$ will result in $S_j$ being undefined. This would yield an experiment that never produces any detection results, and so we remove from consideration the $C_i$ sequences that end with an infinite sequence of $v_{16}$-induced distributions. As we are modeling experiments for which we expect that if we continue running the experiment, we always continue to (eventually) witness new detection events, this restriction is a natural assumption. With this restriction, it turns out that any local distribution for $S_j$ can be modeled as a convex combinations of the (fifteen) induced local distributions, which we denote $\{\vec{v_k}^{**}\}_{k=1}^{15}$. This holds even if the underlying $C_i$ states vary over time. This intuitively plausible result is formulated in the following proposition:

\medskip

\emph{Proposition.}  The space of local distributions for a given $S_j$ is equivalent to the convex hull of the 15 distributions $\{\vec{v_k}^{**}\}_{k=1}^{15}$ induced by the 15 deterministic strategies $\{v_k\}_{k=1}^{15}$. 

\medskip

The proof of this proposition is given at the end of the section. With this result, we can analyze the convex hull of the 15 distributions $\{\vec{v_k}^{**}\}_{k=1}^{15}$ to understand the space of local distributions for $S_j$. This convex hull is a geometrical object known as a \emph{convex polytope}, and the collection of points in a convex polytope can be characterized by a collection of inequalities that must be satisfied. The space of local probability distributions for $S_j$ is thus found to be the collection of probability distributions over the 12 outcomes satisfying the following constraints:
\begin{eqnarray}\label{e:no-signalling}
p(\text{++}a b) + p(\text{+0} ab) &=& p(\text{++}ab') + p(\text{+0}ab')  \\
p(\text{++} a'b) + p(\text{+0} a' b) &=& p(\text{++} a' b') + p(\text{+0} a'b') \label{e:ns2} \\
p(\text{++} a b) + p(\text{0+} a b) &=& p(\text{++} a' b) + p(\text{0+} a' b) \label{e:ns3} \\
p(\text{++} a b') + p(\text{0+}a b') &=& p(\text{++} a' b') + p(\text{0+} a'b') \label{e:ns4}
\end{eqnarray}
\begin{eqnarray}\label{e:E1}
p(\text{++} ab) - p(\text{+0}ab') -p(\text{0+}a'b) -p(\text{++}a'b') &\le& 0  \\
p(\text{++} ab') - p(\text{+0}ab) -p(\text{0+}a'b') -p(\text{++}a'b) &\le& 0 \label{e:Eb2} \\
p(\text{++} a'b) - p(\text{+0}a'b') -p(\text{0+}ab) -p(\text{++}ab') &\le& 0 \label{e:Eb3} \\
p(\text{++} a'b') - p(\text{+0}a'b) -p(\text{0+}ab') -p(\text{++}ab) &\le& 0. \label{e:Eb4}
\end{eqnarray}
Given the equiprobable distribution of the measurement settings, the equations (\ref{e:no-signalling})--(\ref{e:ns4}) are equivalent to no-signaling conditions (obeyed by local and quantum theories alike). The expressions (\ref{e:E1})--(\ref{e:Eb4}) are proper Bell inequalities. Note that (\ref{e:E1})--(\ref{e:Eb4}) are equivalent to each other up to permutation of detector settings, and (\ref{e:E1}) is equivalent to the Eberhard inequality (\ref{e:eberhardo}). 

So, there is essentially one CH-type inequality (the Eberhard version), and all others can be obtained from it by permuting it and/or combining it with one or more of the no-signaling relations. These are robust to variations over time in the state of the system, as the distribution $C_i$ is allowed to vary for each trial in the proof below. The $S_j$ variables can have different distributions from each other, but each must satisfy the constraints (\ref{e:no-signalling})-(\ref{e:Eb4}).

\medskip

We now prove the proposition. We will use the notation $\vec{C_i}$ and $\vec{S_i}$ to denote the probability distributions for the random variables $C_i$ and $S_i$, respectively, and we continue to use the assumptions of the paper: the experiment is performed as a sequence of discrete trials and the measurement settings are governed by two independent random processes that re-assign the measurement settings prior to each experimental trial with probability 50\%. We also have the additional postulate that each $S_j$ is well-defined, taking a value among the 12 possible CH outcomes with probability one -- so a $C_i$ sequence ending with an infinite sequence of $v_{16}$ states is ruled out. The first step in proving the proposition is to prove the following claim.

\medskip

\noindent\emph{Claim 1:} If the experimental assumptions are met, then under locality and the above postulate, the probability distribution for $S_j$ can be expressed as a convex combination of the 15 induced local deterministic strategies $\{\vec{v_k}^{**}\}_{k=1}^{15}$.

\medskip

\noindent\emph{Proof:} For simplicity of exposition, we prove the result for $S_1$, but the same argument will apply to any $S_j$. Let $\vec{C_i^*}$ denote the restriction of $\vec{C_i}$ to the (sub)probability distribution on the elements of  $K$ (the 12 non-00 outcomes), and let $q_i=P(C_i\notin K)$ with $q_0 = 1$. Then the distribution of $S_1$ can be expressed as
\begin{equation}\label{e:S1first}
\vec{S_1} = \sum_{i=1}^\infty \bigg[\prod_{j=0}^{i-1} q_j\bigg]\vec{C_i^*}.
\end{equation}
To understand why (\ref{e:S1first}) holds, consider what happens on the first trial. If $C_1$ assumes a value in the set $K$, then $S_1$ will take this value. The probability of the various possibilities for $S_1$ under this contingency will be given by the elements of $\vec{C_i^*}$. There is a probability $q_1$ that $C_1$ will not take a value in $K$, in which case $S_1$ will remain undefined as we move to the second trial. If $C_2$ then takes a value in the set $K$, $S_1$ will then assume this value, but there was only a probability of $q_1$ of getting to this point without already setting $S_1$, so we only add $q_1*\vec{C_i^*}$ to the probability distribution of $S_1$. Moving to the third trial, we will add $q_1*q_2*\vec{C_i^*}$, etc., and repeating the process out to infinity yields (\ref{e:S1first}). 

Now, for an experiment governed by a local hidden variable theory, Fine's result allows us to express the general distribution for a given $C_i$ as
$$
\vec{C_i} = \sum_{k=1}^{16} c_{i_k}\vec{v_k},
$$
where the $c_{i_k}$ form a collection of nonnegative real numbers whose sum is $1$. So for the infinite sum in (\ref{e:S1first}), some of the $\vec{C_i^*}$ distributions may be the degenerate distribution induced by $v_{16}$, corresponding to $c_{i_{16}}=1$. (In an experimental setting, we indeed expect \emph{most} of the trials to look this way, with no count at either detector.) We would like to remove these terms from the expression (\ref{e:S1first}). Define $T\subseteq\mathbb{N}$ to be $T =\{i| \vec{C_i^*}\ne \vec 0 \}$. Note that if $\vec{C_k^*}$ is the $n$th occurrence of a non-$\vec 0$ distribution, and $\vec{C_l^*}$ comes some trials later with the $(n+1)$th non-$\vec 0$ distribution, then $q_k$ will be less than 1, but $q_{k+1}=q_{k+2}=...=q_{l-2}=q_{l-1}=1$, so the intervening $q$ terms in the product in (\ref{e:S1first}) can be dropped without changing the value of the product. Hence we can rewrite (\ref{e:S1first}) as
\begin{equation}\label{e:S1second}
\vec{S_1} = \sum_{i\in T} \bigg[\prod_{(T\cup\{0\})\cap \{j|j<i\}} q_j\bigg]\vec{C_i^*}.
\end{equation}
As (\ref{e:S1second}) is a little unwieldy, we continue to work from expression (\ref{e:S1first}), taking the indices to have been re-enumerated so as to only to refer to the $\vec{C_i^*}$ for which $i\in T$. If we define
\begin{equation*}
p_{v_k}= \text{the sum of the 12 CH components of $\vec{v_k}$,}
\end{equation*}
we can see that $\vec{C_i^*}$ equals $\sum_{k=1}^{15}c_{i_k}(p_{v_k}\vec{v_k}^{**})$, so we can write
\begin{eqnarray}\label{e:S1third}
\vec{S_1} &=& \sum_{i=1}^\infty \bigg[\prod_{j=0}^{i-1} q_j\bigg]\sum_{k=1}^{15}c_{i_k}p_{v_k}\vec{v_k}^{**} \notag \\ 
&=&\sum_{k=1}^{15}\bigg\{\sum_{i=1}^\infty \bigg[\prod_{j=0}^{i-1} q_j\bigg]c_{i_k}p_{v_k}\bigg\}\vec{v_k}^{**}.
\end{eqnarray}
To prove the proposition, we must show that the expressions in the curly braces in (\ref{e:S1third}) are nonnegative and sum to 1. The nonnegativity is clear. As for summing to 1, we have 
\begin{eqnarray}\label{e:S1fourth}
\sum_{k=1}^{15}\bigg\{\sum_{i=1}^\infty \bigg[\prod_{j=0}^{i-1} q_j\bigg]c_{i_k}p_{v_k}\bigg\} &=& \sum_{i=1}^\infty \bigg[\prod_{j=0}^{i-1} q_j\bigg]\sum_{k=1}^{15}c_{i_k}p_{v_k} \notag \\
&=& \sum_{i=1}^\infty \bigg[\prod_{j=0}^{i-1} q_j\bigg] p_i,
\end{eqnarray}
where $p_i=P(C_i\in K)=1-q_i$. But the expression in (\ref{e:S1fourth}) is equal to exactly the probability that $S_1$ eventually assumes a value -- that is, the probability that if we keep running the trials out to infinity, we eventually get some sort of a detection result other than $00$. Since it is taken as a postulate that this probability equals one, the proof is complete. $\hfill\Box$

\medskip

By the above result, any local strategy for $S_j$ can be expressed as an element of the convex hull of the deterministic strategies $\{\vec{v_k}^{**}\}_{k=1}^{15}$. Furthermore, the converse holds: any convex combination of these 15 deterministic strategies can be realized by a sequence of local distributions $\{\vec{C_i}\}$, and is thus a valid local distribution for $S_j$. In fact, a slightly stronger statement can be proved, as seen below. Combined with Claim 1, the following result implies the main proposition of this section.

\medskip

\noindent\emph{Claim 2:} If the probability distribution $\vec{S_j}$ is in the convex hull of the 15 deterministic strategies $\{\vec{v_k}^{**}\}_{k=1}^{15}$, then there exists a sequence of distributions $\{\vec{C_i}\}$ inducing $\vec{S_j}$ such that the $\vec{C_i}$ are identical, independent, and local. 

\medskip

\noindent\emph{Proof:} We prove the claim for $S_1$, and the argument can apply to all $S_j$. By assumption,
\begin{equation}\label{e:slocal}
\vec{S_1} = \sum_{k=1}^{15}d_k\vec{v_k}^{**},
\end{equation}
where $\{d_k\}$ is a set of nonnegative real numbers whose sum is 1. To come up with a similar expression for $\vec{C_i}$, first define the following constants:
\begin{equation}\label{e:tuvxyzdef}
x = \sum_{k=1}^7 d_k \quad\quad\quad y= \sum_{k=8}^{11} d_k \quad\quad\quad z= \sum_{k=12}^{15} d_k 
$$ $$
s = {3\over 3x+ 4y + 6z}           \quad\quad\quad      t = {4\over 3x+ 4y + 6z}
$$ $$
 u = {6\over 3x+ 4y + 6z}.
\end{equation}
The constants $x$, $y$, and $z$ are chosen to sort and add the $\vec{v_i}^{**}$ coefficients for which the corresponding $p_{v_k}$ values are $1$, ${3\over 4}$, and ${1\over 2}$, respectively. Note that $x+y+z=1$. It turns out that $\{\vec{C_i}\}$ realizes $\vec{S_1}$ if the $C_i$ are independent and identically distributed with the distribution
\begin{equation}\label{e:cithatrealizess1}
\vec{C_i} = \sum_{k=1}^7 s d_k \vec{v_k}+\sum_{k=8}^{11} t d_k \vec{v_k} + \sum_{k=12}^{15} u d_k \vec{v_k}.
\end{equation}
To check that this is true, note that the distribution of $S_1$ can be obtained by conditioning on the set of events, ``the value of $S_1$ is set by the $i$th trial," for $i$ ranging from one to infinity. For each trial, there is a probability $q$ of passing to the next trial without setting the value of $S_1$, so we can write the distribution of $S_1$ induced by (\ref{e:cithatrealizess1}) as
\begin{equation}\label{e:s1infinitesum}
\vec{S_1}_{\text{(induced)}} = \vec{C_1}^* + q\vec{C_2}^*+q^2\vec{C_3}^*+q^3\vec{C_4}^* + \cdots
\end{equation}
where $q={1\over 4}yt+{1\over 2}zu$ is the probability that a given $C_i$ random variable yields one of the four ``00" outcomes. As all of the $\vec{C_i}^*$ are identically distributed with the distribution (\ref{e:cithatrealizess1}), the expression (\ref{e:s1infinitesum}) can be rewritten as 
\begin{widetext}
\begin{eqnarray*}
\vec{S_1}_{\text{(induced)}} &=& {1\overwithdelims() 1-q}\vec{C_i}^* \\
&=& {3x+4y+6z\overwithdelims() 3}\bigg[\sum_{k=1}^7 s d_k \cdot 1 \cdot\vec{v_k}^{**}+\sum_{k=8}^{11} t d_k \cdot{3\overwithdelims () 4}\cdot\vec{v_k}^{**} + \sum_{k=12}^{15} u d_k \cdot{1\overwithdelims()2}\cdot\vec{v_k}^{**}\bigg]\\
&=& \sum_{k=1}^7 d_k\vec{v_k}^{**}+\sum_{k=8}^{11} d_k \vec{v_k}^{**} + \sum_{k=12}^{15} d_k \vec{v_k}^{**}= \sum_{k=1}^{15}d_k\vec{v_k}^{**}. \hspace{7.5cm}\Box\\
\end{eqnarray*}
\end{widetext}

\medskip

\subsection{Effects of Modifying the Setting Probability Distribution}\label{s:mod}

In the paper, it was presumed that the four setting configurations are equiprobable. This is a natural choice for setting probabilities, as it allows for Eberhard-style quantities such as $J$ to be analyzed in a very straightforward manner. However, other setting probabilities can be considered. At a minimum, one should at least consider small deviations from equiprobability, in order to make the experimental model robust. This is desirable because the statement ``the probability that the setting will be $a$ is exactly ${1\over 2}$" is empirically problematic, whereas the statement ``the probability that the setting will be $a$ is within $\epsilon$ of ${1\over 2}$" is easier to justify.

\medskip

\emph{Small deviations from equiprobable settings.} As described in footnote [16], the setting data from the experiment \cite{christensen:2013} exhibited small-but-significant deviations from equiprobability, but any experiment would have trouble making the case that the measurement setting probabilities are ``exactly" 50-50. For example, in \cite{giustina:2013}, equiprobable settings are simulated by operating the detectors for 300 seconds in each of the four configurations, during which an estimated 24.2 M pairs are emitted per time block. Therefore the probability of a particular setting configuration is not exactly 25\%, but depends on the accuracy and precision of the experimenter's clock. Thus it is useful to have a little wiggle room in the mathematical model to allow for small deviations from equiprobability in the measurement settings. 

We return now to the analysis of the statistic $J_k$. Earlier it was assumed that the setting probabilities were 50\%; now, this is replaced with the assumption that 
\begin{eqnarray}\label{e:probsepsilon}
{1\over 2} - \epsilon \le p_a \le {1\over 2} + \epsilon \notag \\
{1\over 2} - \epsilon \le p_b \le {1\over 2} + \epsilon,
\end{eqnarray}
where $p_a$ is defined to be the probability that the setting at detector 1 is $a$, and $p_b$ is the probability that the setting at detector 2 is $b$. We retain the assumption that the two setting processes are independent from each other, but the relationship $E(J_k)\le 0$ is no longer necessarily true. The following claim expresses a new bound for $E(J_k)$.

\medskip

\noindent\emph{Claim.} If the setting probabilities obey the constraints (\ref{e:probsepsilon}), then for any $\epsilon < {1\over 2}$, the following inequality holds:
\begin{equation}\label{e:4ebound}
E(J_k)\le {4\epsilon \over 1 + 4\epsilon^2}.
\end{equation}

\noindent\emph{Proof:} First analyze each of the sixteen possible deterministic distributions $\vec{v_k}$ for $C_i$. Let $q_a=1-p_a$ and $q_b=1-p_b$, and let the event $E_i$ be the condition that $C_i$ takes on one of the four values in the set $E= \{\text{++}ab, \text{+0}ab', \text{0+}a'b,\text{++}a'b' \}$. Starting with $\vec{v_1}$, and referring to Table \ref{t:deterministicstrategies}, one can see that
\begin{multline}\label{e:4eboundv}
 p_{\vec{v_1}}(\text{++}ab|E_i) - p_{\vec{v_1}}(\text{+0}ab'|E_i) \\  -p_{\vec{v_1}}(\text{0+}a'b|E_i) -p_{\vec{v_1}}(\text{++}a'b'|E_i)\\
={P_{\vec{v_1}}(\text{++}ab, E_i) -P_{\vec{v_1}}(\text{++}a'b', E_i) \over P_{\vec{v_1}}(E_i) } \\
= {p_a p_b - q_a q_b\over p_a p_b + q_a q_b}. 
\end{multline}
Calculus can be used to determine the maximum of the expression (\ref{e:4eboundv}) for $p_a, p_b \in [{1\over 2} -\epsilon, {1\over 2} + \epsilon]$. The maximum occurs at $p_a=p_b={1\over 2} + \epsilon$, so long as $\epsilon < {1\over 2}$, and the maximum is ${4\epsilon \over 1+ 4\epsilon^2}$. It is straightforward to similarly examine the other deterministic distributions $\vec{v_k}$, finding that the upper bound ${4\epsilon \over 1+ 4\epsilon^2}$ also holds for the other deterministic distributions $\vec{v_k}$ for which $P_{\vec v_k}(E_i)>0$. (Incidentally, only $\vec{v_1}$ saturates this bound.)

Now, consider a $C_i$ for which $P_{\vec{C_i}}(E_i)>0$. $\vec{C_i}$ can be represented as a convex combination of the $\vec{v_k}$; clearly this representation must assign nonzero weight to at least one of the $\vec{v_k}$ for which $P_{\vec {v_k}}(E_i)>0$. Thus if we define the set $\mathcal{E}=\{k|P_{\vec {v_k}}(E_i)>0\}$, we can write 
\begin{widetext}
\begin{eqnarray}\label{e:4eboundC}
 p_{\vec{C_i}}(\text{++}ab|E_i) &-& p_{\vec{C_i}}(\text{+0}ab'|E_i) -p_{\vec{C_i}}(\text{0+}a'b|E_i) -p_{\vec{C_i}}(\text{++}a'b'|E_i) \notag \\
&=& {P_{\vec{C_i}}(\text{++}ab, E_i) - P_{\vec{C_i}}(\text{+0}ab', E_i) -P_{\vec{C_i}}(\text{0+}a'b, E_i) -P_{\vec{C_i}}(\text{++}a'b', E_i)\over P_{\vec{C_i}}(E_i)} \notag \\
&=& {\sum_{k\in \mathcal{E}}c_{i_k}P_{\vec{v_k}}(\text{++}ab, E_i) - \sum_{k\in \mathcal{E}}c_{i_k}P_{\vec{v_k}}(\text{+0}ab', E_i)- \sum_{k\in \mathcal{E}}c_{i_k}P_{\vec{v_k}}(\text{0+}a'b, E_i)- \sum_{k\in \mathcal{E}}c_{i_k}P_{\vec{v_k}}(\text{++}a'b', E_i) \over  \sum_{k\in \mathcal{E}}c_{i_k}P_{\vec{v_k}}(E_i)} \notag \\
&=& {\sum_{k\in \mathcal{E}}c_{i_k}\big [ p_{\vec{v_k}}(\text{++}ab| E_i) - p_{\vec{v_k}}(\text{+0}ab'| E_i)- p_{\vec{v_k}}(\text{0+}a'b| E_i)- p_{\vec{v_k}}(\text{++}a'b'| E_i)\big ]P_{\vec{v_k}}(E_i) \over  \sum_{k\in \mathcal{E}}c_{i_k}P_{\vec{v_k}}(E_i)} \notag \\
&\le& {4\epsilon \over 1+ 4\epsilon^2}.
\end{eqnarray}
\end{widetext}
Finally, as the $J_k$ random variables take values whenever the root $C_i$ random variables assume outcomes in the set $E$, it follows that that $E(J_k)\le{4\epsilon \over 1+ 4\epsilon^2}$.  \hfill $\Box$

\medskip

As is clear from the proof, the bound (\ref{e:4ebound}) is tight, saturated when $\vec{C_i}= \vec{v_1}$ and $p_a=p_b={1\over 2} + \epsilon$. Whereas before, one could say that $P(J_k=+1)\le {1\over 2}$ under a local theory, now this must be replaced with the weaker constraint 
\begin{equation}\label{e:Jeps}
P(J_k=+1)\le {1\over 2}+ {2\epsilon\over 1+4\epsilon^2}.
\end{equation}
This gives us a quick way to adjust confidence levels if the deviations from equiprobability are not too large. For instance, in \cite{christensen:2013}, about 50.58\% of the trials were measured with setting $b$, which would represent a statistically significant deviation from 50\% given the millions of trials. The proportion of $a$ settings did not stray as far from $50\%$, and so an $\epsilon$ value of .006 could be used. If we form the $J$-type statistic from equation (\ref{e:E2}) we get $J_{(\ref{e:E2})}=2,414$, with $65,876$ total movements of the statistic. The claim of this section still applies to $J_k$ if it is redefined to form $J_{(\ref{e:E2})}$, so appealing to the bound (\ref{e:Jeps}), one can use the binomial distribution with a modified probability of success $.512$ to obtain a still-significant p-value of $.00058$.

\medskip

\emph{Large deviations from equiprobable settings.} The bound (\ref{e:Jeps}) becomes useless if the deviations from equiprobable settings are too large. Instead, the test statistic $J_k$ must be modified to reflect the true probability settings. Updating the definition of $J_k$, we obtain a generalized definition
\begin{equation}\label{e:Jgen}
J_k = \begin{cases} +{1\over p_ap_b}, & \text{if $T_k=$++$ab$ } \\ 
 -{1\over p_aq_b}, & \text{if $T_k=$+0$ab'$} \\
 -{1\over q_ap_b}, & \text{if $T_k=$0+$a'b$} \\
 -{1\over q_aq_b}, & \text{if $T_k=$++$a'b'$}, 
\end{cases}
\end{equation}
where we recall that $T_k$ is the outcome of the $k$th trial that takes a value in the set $E$. Working with the generalized $J_k$ complicates the analysis somewhat, but it is possible that some benefit could be gained by changing the setting probabilities. Though the minimum detector efficiency required to demonstrate non-locality cannot be lowered, it is possible that a different distribution over the setting probabilities may lower the number of trials needed to demonstrate non-locality. However, as noted in \cite{gillvandam:2005}, ``to find the (joint) setting distribution that optimizes the [statistical] strength of a non-locality proof is a highly nontrivial computation," and the scenario of this paper is no exception to this rule. 

To begin the analysis, one can show that $E(J_k)\le 0$ still must hold for $J_k$ as defined in (\ref{e:Jgen}). Unfortunately, the general $J_k$ can assume 3 or 4 values instead of just 2, so the induced random walk $J$ must be analyzed with the more-difficult non-binary martingale methods, as was previously done with the statistic $Ch$. The random walk $J$ also is no longer restricted to the integers, though at every step there are only finitely many values it can take. In such an experiment, one can still use the back-tracing method described in Section \ref{s:statistics} to find exact values for statistical strength. For a computationally simpler starting point, one can apply an Azuma-Hoeffding inequality to get a quick upper bound as a first step for exploring the statistical effects (possibly beneficial) of varied setting probabilities.

\end{document}